\documentclass[pt11,aps,superscriptaddress,notitlepage]{revtex4-1}%
\usepackage{amsmath}
\usepackage{amssymb}
\usepackage{amsfonts}
\usepackage{revsymb}
\usepackage{graphicx}
\usepackage{color}
\usepackage{hyperref}%
\providecommand{\U}[1]{\protect\rule{.1in}{.1in}}

\newcommand{\ket}[1]{{|#1\rangle}}

\begin{document}

\title{Loss of Information in Quantum Guessing Game}
\begin{abstract}
Incompatibility of certain measurements -- impossibility of obtaining deterministic outcomes simultaneously -- is a well known property of quantum mechanics.
This feature can be utilized in many
contexts, ranging from Bell inequalities to device dependent QKD protocols. Typically, in these applications the measurements are chosen from a predetermined set based on a classical
random variable. One can naturally ask, whether the non-determinism of the outcomes is due to intrinsic hiding property of quantum mechanics, or rather by the fact that classical, incoherent information entered the system via the choice of the measurement. Authors of \cite{Rozpedek2017} examined this question for a specific case of two mutually unbiased measurements on systems of different dimensions. They have somewhat surprisingly shown that in case of qubits, if the measurements are chosen coherently with the use of a controlled unitary, outcomes of both measurements can be guessed deterministically.
Here we extend their analysis and
show that specifically for qubits, measurement result for any set of measurements with any a-priori probability distribution can be faithfully guessed by a suitable state preparation and measurement. We also show that up to a small set of specific cases, this is not possible for higher dimensions. This result manifests a deep difference in properties of qubits and higher dimensional systems and suggests that these systems might offer higher security in specific cryptographic protocols. More fundamentally, the results show that the impossibility of predicting a result of a measurement is not caused solely by a loss of coherence between the choice of the measurement and the guessing procedure.

\end{abstract}

\author{Martin Plesch\footnote{martin.plesch@savba.sk}}

\affiliation{Institute of Computer Science, Masaryk University,  Botanick\'a 68a, 602 00 Brno, Czech Republic}
\affiliation{Institute of Physics, Slovak Academy of Sciences, Bratislava, Slovakia}

\author{Matej Pivoluska}

\affiliation{Institute of Computer Science, Masaryk University,  Botanick\'a 68a, 602 00 Brno, Czech Republic}
\affiliation{Institute of Physics, Slovak Academy of Sciences, Bratislava, Slovakia}
\affiliation{Institute for Quantum Optics and Quantum Information, Vienna}

\maketitle
\section{Introduction}

The impossibility of performing a general set of measurements on a fixed
quantum state without disturbing it lies in the heart of quantum theory.
This concept is often referred to as the measurement problem \cite{Nielsen:2011:QCQ:1972505}.
More concretely, one is not able
to perform subsequently a series of incompatible measurements that would
attain a well defined result, irrespectively on the measured state. A well known consequence of measurement incompatibility is the Heisenberg uncertainty principle \cite{Heisenberg1927,Kennard1927}
which shows the impossibility to specify both position and momentum of a particle
with arbitrary precision. Since this pioneering work, measurement incompatibility has been extensively studied in many contexts,
e.g. uncertainty relations \cite{PhysRevA.90.062127,Berta2014,Christandl2005,Berta2010,Coles2012,Coles2011,Dupuis2015,Frank2013,Furrer2014,
Hall1995,Liu2015,Korzekwa2014,Luo2005,Renes2009,Sanchez-Ruiz1995} or joint measurability \cite{Ludwig1964,Busch1984,Busch1986,Muynck1990,Lahti2003,Wolf2009,Reeb2013,Uola2014,Heinosaari2014,Heinosaari2015}.

The task can be slightly reformulated and instead of performing a  series of
subsequent measurements on the same system, the experimentalist performs only a single measurement on a state.
The question then is, whether such measurement can produce a predictable outcome. The answer here is naturally affirmative, if the experimentalist  is also allowed to choose the measured state,
as for each projective measurement, its eigenstates produce a deterministic outcome. The experiment becomes equivalent to the subsequent measurement experiment,
if instead of using a single measurement, the experimentalist is tasked to perform a randomly chosen measurement out of a predefined set and the goal is to prepare a state which produces
deterministic outcomes for all measurements from this set.

With this modification the task can be conveniently rephrased in terms of a guessing game \cite{Berta2010}.
The players,
Alice and Bob, agree beforehand on a set of measurements that Alice can perform in her lab.
Bob then prepares a probe state and sends it to Alice, who randomly performs one of
the available measurements. The question of interest is to identify the conditions under which Bob is
be able to predict the outcome of all Alice's measurements with certainty.

Clearly, without any information about the choice of Alice, Bob only can guess correctly with probability $1$,
 if there exists a state that is an eigenstate of all measurements in
Alice's set and additionally, all measurements associate the same outcome to this state.
Only little changes when Bob, after the measurement, receives from Alice
classical information about her measurement choice -- here the set of measurements still
needs to share a single common eigenstate, but it does not need to be
associated with the same measurement outcome anymore. Bob's strategy consists of
preparing the common eigenstate as the probe state in the second step of the guessing game.
By learning the specific
measurement performed by Alice, Bob guesses the value associated to the probe eigenstate by Alice's measurement.

One can ask whether the incapability of Bob to learn the
measurement result of Alice in other non-degenerate cases, unique to quantum realm, is caused by intrinsic properties
of the quantum theory or by the fact that Bob
receives ``only'' classical, rather than quantum, information from Alice.

To tackle with this question, one can re-design the guessing game to a fully
quantum level as follows. Instead of letting Alice to choose the measurement randomly on her own,
the choice of the measurement is made coherently with the help of a controlled unitary operation.
After performing the measurement, Alice will send the control quantum system to Bob, sharing
with him the information about her choice on quantum level. Now the
natural question is to what extent this can help Bob determine Alice's measurement outcomes.

Filip Rozpedek \textit{et.~al.} in \cite{Rozpedek2017} examined this question for a specific choice of
two distinct measurements performed by Alice in two Mutually Unbiased Bases
(MUBs). They have shown that specifically for qubit, Bob can learn with
certainty the outcome of the measurement by sending a specifically designed
probe state. They have also shown that for higher dimensions and two fixed
measurements in MUBs, perfect guessing is not possible anymore. This was
intuitively explained by the fact that in the latter case, it is naturally
impossible to perfectly determine a measurement on a higher dimensional system
by measuring a qubit.

In this paper we extend the analysis of this game in two main directions. First, we fully
characterize the game with qubit measurements. We show that Bob is always able to predict the
measurement result of Alice, even if Alice is able to choose measurements from an \emph{arbitrary large} set,
with the assumption that he knows the set of measurements Alice is using
and the underlying probability distribution Alice uses to choose them in the experiment.
This basically means that quantum regime with qubits is very similar to the classical scenario.

In the second part we analyze the guessing game with measurements on systems of higher dimensions.
By parameter analysis we show that for larger quantum systems, Bob is not able to learn the measurement result of Alice with
certainty, except for some degenerate cases. This is particularly also true in cases, where Alice uses the same or higher number of
measurements compared to the dimension of Bob's system.
In other words, Bob cannot correctly guess the measurement outcome even in cases where he receives a quantum system of dimension
higher than the number of possible measurement outcomes.
This result therefore undermines the intuition that Bob's incapability of guessing the correct outcome is caused  by the deficient dimension
of the system he receives from Alice.

The paper is organized as follows. In chapter \ref{chap2} we rigorously define
the guessing game. In chapter \ref{chap3} we first analyze the qubit
system under uniform selection of a measurement from an arbitrary set and subsequently
we extend our analysis to non-uniform situations as well. In chapter
\ref{chap4} we use parameter counting analysis to show that perfect guessing is only possible for
the set of measurements of {measure zero}. Particularly, we explicitly show that for
a set of three mutually unbiased measurements on a qutrit the perfect guessing is not possible.
We conclude in chapter \ref{chap5}.

\section{The guessing game}\label{chap2}

\begin{figure}
\includegraphics[scale = 2]{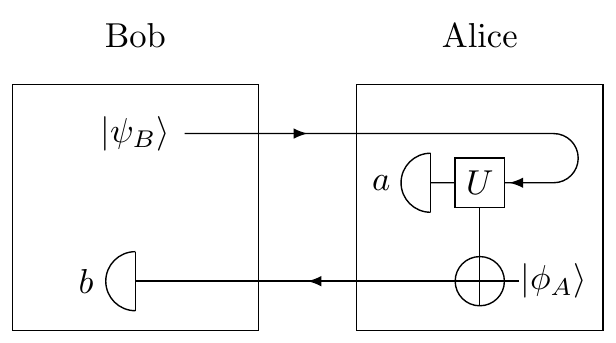}
\caption{\emph{The guessing game.} Bob prepares a probe state $\ket{\psi_B}$ and sends it to Alice. Alice prepares a pure state $\ket{\phi_A}$, which is in conjunction with a controlled unitary $U$ used to implement different projective measurements on Bob's state. The measurement outcome of Alice's measurement is denoted $a$. Subsequently the control state is sent to Bob, who measures it in order to obtain a classical measurement outcome $b$. Bob's goal is to prepare a probe state such that the probability of $a=b$ is as high as possible, in the ideal case equal to $1$.}
\label{fig:GuessingGame}
\end{figure}

The guessing game is schematically depicted in FIG. \ref{fig:GuessingGame}. Alice and Bob
agree on a dimension $B$ and a set of $A$ projective measurements $M_{i}$ in
this dimension. Bob prepares a state $\left\vert \psi_{B}\right\rangle $ of
dimension $B$ and sends the state to Alice. Alice prepares a state
\begin{equation}
\left\vert \phi_{A}\right\rangle =\frac{1}{\sqrt{A}}%
{\displaystyle\sum\limits_{i=0}^{A-1}}
\left\vert i\right\rangle \label{fi_A}%
\end{equation}
and a set of unitary operations $U^{\dagger}_{i}$ acting on dimension $B$ such that%
\begin{equation}
U_{i}\left\vert j\right\rangle =\left\vert M_{i}\right\rangle _{j},\label{U_i}%
\end{equation}
where $\left\vert M_{i}\right\rangle _{j}$ is the state that yields an outcome $j$, if
$i$-th measurement is performed. Alice now performs a controlled operation
on both states
\begin{equation}
U\left\vert \phi_{A}\right\rangle \left\vert \psi_{B}\right\rangle =\frac
{1}{\sqrt{A}}%
{\displaystyle\sum\limits_{i=0}^{A-1}}
\left\vert i\right\rangle U_{i}^{\dag}\left\vert \psi_{B}\right\rangle
\label{U_cond}%
\end{equation}
and performs a measurement in the computational basis on $\left\vert \psi
_{B}\right\rangle $, obtaining a result $a$ of dimension $B$. In the next step,
Alice sends the state $\left\vert \phi_{A}\right\rangle $ to Bob and asks him to
guess her measurement outcome. In order to produce a correct guess, Bob measures
the state and produces an outcome $b$.

One can generalize the guessing game by allowing Alice to prepare a more
general control state in the form
\begin{equation}
\left\vert \phi^{\zeta}_{A}\right\rangle =%
{\displaystyle\sum\limits_{i=0}^{A-1}}
\zeta_{i}\left\vert i\right\rangle .\label{fi_niid}%
\end{equation}
Now one can analyze different scenarios.
In the first one, the amplitudes $\zeta_{i}$ are fully known to Bob.
On the other side of the spectrum, Bob doesn't have any information about $\zeta_i$
 and in the intermediate scenarios Bob can have some partial information
about $\zeta_i$. It is easy to see
that complete ignorance of $\zeta_{i}$ will not allow Bob to guess any better
than in the classical scenario. In order to see this, consider a set of two incompatible
measurements without a common eigenstate. If one of the measurements is
chosen deterministically (say $\zeta_{0}=1$), Bob has to choose one of the
eigenstates $\left\vert M_{0}\right\rangle _{j}$, in order to guess Alice's outcome with probability $1$. On the other hand, for
$\zeta_{1}=1$ Bob has to choose one of the eigenstates $\left\vert M_{1}\right\rangle _{j}$.
Therefore if two measurements do not share an eigenstate and Bob has no information about $\zeta_{0}$, he
cannot guess Alice's outcome with certainty. In what follows, we analyze
the scenarios in which Bob has at least some partial knowledge of $\zeta_{i}$.

\section{Guessing game with qubits}\label{chap3}

In this section we analyze the probability of correct guessing in scenarios where $\ket{\psi_B}$ is a qubit.
In this case $\ket{\psi_B}$ can be parameterized as
\begin{equation}
\left\vert \psi\right\rangle =\alpha\left\vert 0\right\rangle +\beta\left\vert
1\right\rangle ,\label{psi_Bob}%
\end{equation}
where we choose $\alpha$ to be real and positive. Let us first consider
a scenario, where all the measurements are chosen with uniform probability.
In such a case, Alice has a general state
(\ref{fi_A}) and the set of conditional operations $U_{i}$ (\ref{U_i}) for each
$i$ is defined in a following way:%
\begin{align*}
\left\vert 0\right\rangle  &  \rightarrow\left\vert \psi_{i}\right\rangle
=\alpha_{i}\left\vert 0\right\rangle +\beta_{i}\left\vert 1\right\rangle \\
\left\vert 1\right\rangle  &  \rightarrow\left\vert \psi_{i}^{\bot
}\right\rangle =\beta_{i}^{\ast}\left\vert 0\right\rangle -\alpha
_{i}\left\vert 1\right\rangle ,
\end{align*}
where again parameters $\alpha_{i}$ are chosen to be real and positive and $\ast$ denotes
complex conjugation. It is straightforward to see that with suitable
parameters $\alpha_{i}$ and $\beta_{i}$ one can encode any projective measurement on a
qubit into a measurement in the computational basis.

Now we can write down the common state of Alice and Bob after the conditional
operation (\ref{U_cond}) is performed%
\begin{equation}
\left\vert \Psi\right\rangle =\frac{1}{\sqrt{A}}%
{\displaystyle\sum\limits_{i=0}^{A-1}}
\left[  \left(  \alpha\alpha_{i}+\beta\beta_{i}^{\ast}\right)  \left\vert
i\right\rangle \left\vert 0\right\rangle +\left(  \alpha\beta_{i}-\beta
\alpha_{i}\right)  \left\vert i\right\rangle \left\vert 1\right\rangle
\right]  . \label{psi}%
\end{equation}
Then Alice performs a measurement in computational basis and depending on her
outcome, the state that is sent to Bob is
\begin{equation}
\left\vert \phi_{0}\right\rangle =\frac{1}{\sqrt{A}}%
{\displaystyle\sum\limits_{i=0}^{A-1}}
\left(  \alpha\alpha_{i}+\beta\beta_{i}^{\ast}\right)  \left\vert
i\right\rangle \label{fi0}%
\end{equation}
or
\begin{equation}
\left\vert \phi_{1}\right\rangle =\frac{1}{\sqrt{A}}%
{\displaystyle\sum\limits_{i=0}^{A-1}}
\left(  \alpha\beta_{i}-\beta\alpha_{i}\right)  \left\vert i\right\rangle .
\label{fi1}%
\end{equation}
Bob can learn the result of the measurement performed by Alice if and only if
he can distinguish these two states, namely if
\begin{equation}
\langle\phi_{0}\left\vert \phi_{1}\right\rangle =0. \label{perp}%
\end{equation}
Here we remind that all the parameters in (\ref{fi0}) and (\ref{fi1}) are
known to Bob, so he knows the full specification of the states he is trying to
distinguish. This is why (\ref{perp}) is a sufficient condition for Bob's perfect guessing.

Condition (\ref{perp}) can be expressed as a complex equation
\begin{equation}
\frac{1}{A}%
{\displaystyle\sum\limits_{i=0}^{A-1}}
\left(  \alpha\alpha_{i}+\beta\beta_{i}^{\ast}\right)  ^{\ast}\left(
\alpha\beta_{i}-\beta\alpha_{i}\right)  =0, \label{orthogonality}%
\end{equation}
where $\alpha$ and $\beta$ are variables that can be chosen by Bob and parameters
$\alpha_{i}$ and $\beta_{i}$ are chosen by Alice, but known to
Bob. It remains to investigate under which conditions on $\alpha_{i}$ and $\beta_{i}$
the equation (\ref{orthogonality}) has a solution.

In what follows we rewrite (\ref{orthogonality}) into a pair of
equations involving only real parameters. To do so we define
\begin{align}
a &  =\alpha\nonumber\\
a_{i} &  =\alpha_{i}\nonumber\\
b &  =\left\vert \beta\right\vert \nonumber\\
b_{i} &  =\left\vert \beta_{i}\right\vert \nonumber\\
\varphi &  =\arg\left(  \beta\right)  \nonumber\\
\varphi_{i} &  =\arg\left(  \beta_{i}\right)  .\label{Reparam}%
\end{align}
Using this new set of parameters, equation (\ref{orthogonality}) gets a form
\begin{align}%
{\displaystyle\sum\limits_{i}}
\left[  aa_{i}+bb_{i}\left(  \cos\varphi\cos\varphi_{i}+\sin\varphi\sin
\varphi_{i}\right)  -jbb_{i}\left(  \sin\varphi\cos\varphi_{i}-\cos\varphi
\sin\varphi_{i}\right)  \right]&\times
\\\nonumber
\left[  ab_{i}\cos\varphi_{i}-a_{i}%
b\cos\varphi+j\left(  ab_{i}\sin\varphi_{i}-a_{i}b\sin\varphi\right)  \right]
&=0,
\end{align}
where $j=\sqrt{-1}$. We can further separate it into two equations for real
and imaginary part%
\begin{align}%
\nonumber
{\displaystyle\sum\limits_{i}}
\left[  \left(  aa_{i}+bb_{i}\left(  \cos\varphi\cos\varphi_{i}+\sin
\varphi\sin\varphi_{i}\right)  \right)  \left(  ab_{i}\cos\varphi_{i}%
-a_{i}b\cos\varphi\right)
 \right.&+\\ \left.
bb_{i}\left(  \sin\varphi\cos\varphi_{i}%
-\cos\varphi\sin\varphi_{i}\right)  \left(  ab_{i}\sin\varphi_{i}-a_{i}%
b\sin\varphi\right)  \right]   &  =0\label{Real}\\%
\nonumber
{\displaystyle\sum\limits_{i}}
\left[  \left(  aa_{i}+bb_{i}\left(  \cos\varphi\cos\varphi_{i}+\sin
\varphi\sin\varphi_{i}\right)  \right)  \left(  ab_{i}\sin\varphi_{i}%
-a_{i}b\sin\varphi\right)
\right.&-\\ \left.
bb_{i}\left(  \sin\varphi\cos\varphi_{i}%
-\cos\varphi\sin\varphi_{i}\right)  \left(  ab_{i}\cos\varphi_{i}-a_{i}%
b\cos\varphi\right)  \right]   &  =0.\label{Imag}%
\end{align}
Let us now rewrite (\ref{Real}) and (\ref{Imag}) into a form%
\begin{align}
X_{r}b^{2}+Y_{r}ab-X_{r}a^{2} &  =0\label{Real_substituted}\\
X_{I}b^{2}+Y_{I}ab-X_{I}a^{2} &  =0,\label{Imag_substituted}%
\end{align}
where
\begin{align*}
X_{r} &  =%
{\displaystyle\sum\limits_{i}}
a_{i}b_{i}\left(  -\cos^{2}\varphi\cos\varphi_{i}-\cos\varphi\sin\varphi
\sin\varphi_{i}-\sin^{2}\varphi\cos\varphi_{i}+\sin\varphi\cos\varphi
\sin\varphi_{i}\right)  =-%
{\displaystyle\sum\limits_{i}}
a_{i}b_{i}\cos\varphi_{i}\\
Y_{r} &  =%
{\displaystyle\sum\limits_{i}}
\left[  -a_{i}^{2}\cos\varphi+b_{i}^{2}\cos\varphi\cos^{2}\varphi_{i}%
+2b_{i}^{2}\sin\varphi\cos\varphi_{i}\sin\varphi_{i}-b_{i}^{2}\cos\varphi
\sin^{2}\varphi_{i}\right]  \\
Z_{r} &  =%
{\displaystyle\sum\limits_{i}}
a_{i}b_{i}\cos\varphi_{i}=-X_{r}.
\end{align*}
and%
\begin{align*}
X_{I} &  =%
{\displaystyle\sum\limits_{i}}
a_{i}b_{i}\left(  -\cos\varphi\cos\varphi_{i}\sin\varphi-\sin^{2}\varphi
\sin\varphi_{i}+\sin\varphi\cos\varphi_{i}\cos\varphi-\cos^{2}\varphi
\sin\varphi_{i}\right)  =-%
{\displaystyle\sum\limits_{i}}
a_{i}b_{i}\sin\varphi_{i}\\
Y_{I} &  =%
{\displaystyle\sum\limits_{i}}
\left[  -a_{i}^{2}\sin\varphi+2b_{i}^{2}\cos\varphi\cos\varphi_{i}\sin
\varphi_{i}+b_{i}^{2}\sin\varphi\sin^{2}\varphi_{i}-b_{i}^{2}\sin\varphi
\cos^{2}\varphi_{i}\right]  \\
Z_{I} &  =%
{\displaystyle\sum\limits_{i}}
a_{i}b_{i}\sin\varphi_{i}=-X_{I}.
\end{align*}
It is important to observe that $X_{r}=-Z_{r}$ and $X_{I}=Z_{I}$ depend only
on parameters $a_i, b_i$ and $\varphi_i$ and not on variables $\varphi$ and $a$.

The set of equations (\ref{Real_substituted},\ref{Imag_substituted}) can be
joined by multiplying (\ref{Real_substituted}) by $X_I$ and (\ref{Imag_substituted}) by $X_r$ and subtracting them:
\begin{equation}
ba\left(  Y_{r}X_{I}-Y_{I}X_{r}\right)  =0. \label{Joined}%
\end{equation}
This equation can have a solution in three cases. The obvious two are $a=0$
and $b=0$. These imply $b=1$ (or $a=1$ respectively) and thus $X_{r}%
=X_{I}=0$, which corresponds to a very specific set of parameters of the
measurements. Naturally for $a=0$ or $b=0$ the value of the complex phase
$\varphi$ has no physical meaning.

The interesting case is $Y_{r}X_{I}-Y_{I}X_{r}=0$, which translates to
\begin{align}
&  X_{I}%
{\displaystyle\sum\limits_{i}}
\left[  -a_{i}^{2}\cos\varphi+b_{i}^{2}\cos\varphi\cos^{2}\varphi_{i}%
+2b_{i}^{2}\sin\varphi\cos\varphi_{i}\sin\varphi_{i}-b_{i}^{2}\cos\varphi
\sin^{2}\varphi_{i}\right]  =\\
&  =X_{r}%
{\displaystyle\sum\limits_{i}}
\left[  -a_{i}^{2}\sin\varphi+2b_{i}^{2}\cos\varphi\cos\varphi_{i}\sin
\varphi_{i}+b_{i}^{2}\sin\varphi\sin^{2}\varphi_{i}-b_{i}^{2}\sin\varphi
\cos^{2}\varphi_{i}\right]  .\nonumber
\end{align}
We can further restructure the equation to
\begin{align}
&  \cos\varphi%
{\displaystyle\sum\limits_{i}}
\left[  -a_{i}^{2}X_{I}+b_{i}^{2}\cos^{2}\varphi_{i}X_{I}-b_{i}^{2}\sin
^{2}\varphi_{i}X_{I}-2b_{i}^{2}\cos\varphi_{i}\sin\varphi_{i}X_{r}\right]  =\\
&  =\sin\varphi%
{\displaystyle\sum\limits_{i}}
\left[  -a_{i}^{2}X_{r}+b_{i}^{2}\sin^{2}\varphi_{i}X_{r}-b_{i}^{2}\cos
^{2}\varphi_{i}X_{r}-2b_{i}^{2}\cos\varphi_{i}\sin\varphi_{i}X_{I}\right]
.\nonumber
\end{align}
and define the argument
\begin{equation}
\varphi=\arctan\frac{%
{\displaystyle\sum\limits_{i}}
\left[  -a_{i}^{2}X_{I}+b_{i}^{2}\cos^{2}\varphi_{i}X_{I}-b_{i}^{2}\sin
^{2}\varphi_{i}X_{I}-2b_{i}^{2}\cos\varphi_{i}\sin\varphi_{i}X_{r}\right]  }{%
{\displaystyle\sum\limits_{i}}
\left[  -a_{i}^{2}X_{r}+b_{i}^{2}\sin^{2}\varphi_{i}X_{r}-b_{i}^{2}\cos
^{2}\varphi_{i}X_{r}-2b_{i}^{2}\cos\varphi_{i}\sin\varphi_{i}X_{I}\right]  }.
\label{argument}%
\end{equation}
For the specific case of the denominator of (\ref{argument}) being equal to
$0$ we define $\varphi=\frac{\pi}{2}$.

Now the question is whether, substituting the result of (\ref{argument}) into
(\ref{Real_substituted}) there exists a solution for $b$. It is a
quadratic equation with the discriminant
\[
D=a^{2}Y_{r}^{2}+4a^{2}X_{r}^{2},
\]
which is clearly non-negative. The solution for $b$ will then be
\begin{equation}
b=a\frac{\sqrt{Y_{r}^{2}+4X_{r}^{2}}-Y_{r}}{2X_{r}}, \label{b}%
\end{equation}
which still depends on $a$. Imposing the normalization condition we get
\begin{equation}
a=\frac{\left\vert 2X_{r}\right\vert }{\sqrt{8X_{r}^{2}+2Y_{r}^{2}-2Y_{r}%
\sqrt{Y_{r}^{2}+4X_{r}^{2}}}}. \label{a}%
\end{equation}

Thus we can conclude that for every set of measurements that Alice chooses,
Bob can prepare a test state specified by parameters $\varphi$ (\ref{argument}%
) and $a$ (\ref{a}), such that with a suitable measurement in the basis
specified by the states (\ref{fi0}) and (\ref{fi1}) he can predict with
certainty the outcome of the measurement performed by Alice.

\subsection{Non-uniform extension}

Here we extend the analysis to a more general situation where Alice 
decides about the measurements based on a more general state (\ref{fi_niid}).
This includes the situations reachable classically, where Alice would simply
favor some of the measurements comparing to some others, but Alice could also
do a more subtle change by just tweaking the phases between different
measurements.

Two obvious questions arise here: is Bob able to retain his capability of
perfect guessing even for this more complicated situation? And, to what extend
does he need to adapt his strategy based on the knowledge of the state used by Alice?

If Alice uses the state (\ref{fi_niid}) instead of (\ref{fi_A}), it causes the
change of (\ref{psi}) into%
\begin{equation}
\left\vert \Psi_{\zeta}\right\rangle =%
{\displaystyle\sum\limits_{i=0}^{A-1}}
\left[  \zeta_{i}\left(  \alpha\alpha_{i}+\beta\beta_{i}^{\ast}\right)
\left\vert i\right\rangle \left\vert 0\right\rangle +\zeta_{i}\left(
\alpha\beta_{i}-\beta\alpha_{i}\right)  \left\vert i\right\rangle \left\vert
1\right\rangle \right]  \label{psi_niid}%
\end{equation}
and the post-selected states depending on Alice's measurement outcome to
\begin{equation}
\left\vert \phi_{\zeta}^{0}\right\rangle =%
{\displaystyle\sum\limits_{i=0}^{A-1}}
\zeta_{i}\left(  \alpha\alpha_{i}+\beta\beta_{i}^{\ast}\right)  \left\vert
i\right\rangle \label{fi0_niid}%
\end{equation}
and
\begin{equation}
\left\vert \phi_{\zeta}^{1}\right\rangle =%
{\displaystyle\sum\limits_{i=0}^{A-1}}
\zeta_{i}\left(  \alpha\beta_{i}-\beta\alpha_{i}\right)  \left\vert
i\right\rangle .\label{fi1_niid}%
\end{equation}
The orthogonality condition then reads%
\begin{equation}%
{\langle\phi_\zeta^0\vert\phi_\zeta^1\rangle =\displaystyle\sum\limits_{i=0}^{A-1}}
\left\vert \zeta_{i}\right\vert ^{2}\left(  \alpha\alpha_{i}+\beta\beta
_{i}^{\ast}\right)  ^{\ast}\left(  \alpha\beta_{i}-\beta\alpha_{i}\right)
=0.\label{orto_niid}%
\end{equation}

It is straightforward to see that only the absolute values $\left\vert
\zeta_{i}\right\vert ^{2}$ enter in further calculations for orthogonality,
but the phases enter into the specification of the post-selected states
(\ref{fi0_niid}) and (\ref{fi1_niid}). We can further rewrite (\ref{orto_niid}%
) to
\begin{equation}%
{\displaystyle\sum\limits_{i=0}^{A-1}}
\left(  \alpha\left\vert \zeta_{i}\right\vert \alpha_{i}+\beta\left\vert
\zeta_{i}\right\vert \beta_{i}^{\ast}\right)  ^{\ast}\left(  \alpha\left\vert
\zeta_{i}\right\vert \beta_{i}-\beta\left\vert \zeta_{i}\right\vert \alpha
_{i}\right)  =0\label{orto_niid2}%
\end{equation}
and redefine the parameters used in the previous subsection (\ref{Reparam})
into%
\begin{align}
a &  =\alpha\nonumber\\
a_{i} &  =\left\vert \zeta_{i}\right\vert \alpha_{i}\nonumber\\
b &  =\left\vert \beta\right\vert \nonumber\\
b_{i} &  =\left\vert \zeta_{i}\right\vert \left\vert \beta_{i}\right\vert
\nonumber\\
\varphi &  =\arg\left(  \beta\right)  \nonumber\\
\varphi_{i} &  =\arg\left(  \beta_{i}\right)  .\label{Reparam_niid}%
\end{align}
Now it is straightforward to see that the specification of the optimal probe
state defined by (\ref{a}, \ref{b}, \ref{argument}) combined with
(\ref{Reparam_niid}) is valid for the more general nonuniform case as well.
Moreover, the definition of the probe state only depends on the absolute
values of the parameters $\zeta_{i}$, meaning Bob only needs the information
about the frequencies of measurements beforehand. On the other hand, phases of
$\zeta_{i}$ enter into the specification of the two post-selected states that
Bob needs to distinguish. So before performing his measurement, Bob needs to
know the phases as well.

\section{Measurements on a larger system}\label{chap4}

It was recently shown in \cite{Rozpedek2017} that for a specific case of two unbiased
measurements, perfect guessing of their outcomes on a system larger then a
qubit is not possible. From one point of view it sounds natural that with more
possible measurement results the outcome will be harder to guess. Especially
if $B>A$, the amount of information extractable in principle from the state
Bob gets from Alice is smaller than the amount Bob needs for perfect guessing.
On the other hand, by increasing the number of possible measurements $A$, Bob
gets a larger system and, at least in principle, it could be easier for him to
learn the outcome of the measurement. Let us now analyze in detail the
simplest non-trivial configuration, namely the case of three unbiased
measurements on a qutrit system.

\subsection{Three unbiased measurements on a qutrit}

Similarly as in the previous section, we characterize the state prepared by Bob
by
\begin{equation}
\left\vert \psi\right\rangle =\alpha\left\vert 0\right\rangle +\beta\left\vert
1\right\rangle +\gamma\left\vert 2\right\rangle .\label{qutrit_Bob}%
\end{equation}
Alice shall perform measurements in one of mutually unbiased bases. These are
generally defined for prime $d$ by
\begin{equation}
\left\vert M_{k}\right\rangle _{i}=\frac{1}{\sqrt{d}}\sum_{j=0}^{d-1}%
\omega^{kj^{2}+ij}\left\vert j\right\rangle ,\label{Mk}%
\end{equation}
where $d$ is the dimension of the system, $k$ is the index of the base
(starting from $0$) and $\omega=e^{\left(  \frac{2\pi i}{d}\right)}$ \cite{WOOTTERS1989363}. These
bases can be obtained from the computational basis by applying a unitary
transformation defined element-wise by
\begin{equation}
U_{k}[i,j]=\frac{1}{\sqrt{d}}\omega^{kj^{2}+ij}.\label{Uk}%
\end{equation}
We take (\ref{Uk}) as the unitary operations associated with measurements
(\ref{U_i}) and build up the conditional operation (\ref{U_cond}).

For the specific case of a qutrit, the operations performed by Alice before
measurement are
\begin{equation}
U_{0}^{\dag}=\frac{1}{\sqrt{3}}\left(
\begin{array}
[c]{ccc}%
1 & 1 & 1\\
1 & \omega^{2} & \omega\\
1 & \omega & \omega^{2}%
\end{array}
\right)  ,U_{1}^{\dag}=\frac{1}{\sqrt{3}}\left(
\begin{array}
[c]{ccc}%
1 & \omega^{2} & \omega^{2}\\
1 & \omega & 1\\
1 & 1 & \omega
\end{array}
\right)  ,U_{2}^{\dag}=\frac{1}{\sqrt{3}}\left(
\begin{array}
[c]{ccc}%
1 & \omega & \omega\\
1 & 1 & \omega^{2}\\
1 & \omega^{2} & 1
\end{array}
\right)  ,\label{qutrit_MUBs}%
\end{equation}
where $\omega=e^{\left(  \frac{2\pi i}{3}\right)}  $. The common state of Alice
and Bob after performing conditional operation by Alice and before the
measurement reads
\begin{align}
\left\vert \Psi\right\rangle  &  =\frac{1}{\sqrt{3}}\left\vert 0\right\rangle
\left(  \alpha\left\vert 0\right\rangle +\alpha\left\vert 1\right\rangle
+\alpha\left\vert 2\right\rangle +\beta\left\vert 0\right\rangle +\omega
^{2}\beta\left\vert 1\right\rangle +\omega\beta\left\vert 2\right\rangle
+\gamma\left\vert 0\right\rangle +\omega\gamma\left\vert 1\right\rangle
+\omega^{2}\gamma\left\vert 2\right\rangle \right)  +\label{psi_qutrit}\\
&  +\frac{1}{\sqrt{3}}\left\vert 1\right\rangle \left(  \alpha\left\vert
0\right\rangle +\alpha\left\vert 1\right\rangle +\alpha\left\vert
2\right\rangle +\omega^{2}\beta\left\vert 0\right\rangle +\omega
\beta\left\vert 1\right\rangle +\beta\left\vert 2\right\rangle +\omega
^{2}\gamma\left\vert 0\right\rangle +\gamma\left\vert 1\right\rangle
+\omega\gamma\left\vert 2\right\rangle \right)  \nonumber\\
&  +\frac{1}{\sqrt{3}}\left\vert 2\right\rangle \left(  \alpha\left\vert
0\right\rangle +\alpha\left\vert 1\right\rangle +\alpha\left\vert
2\right\rangle +\omega\beta\left\vert 0\right\rangle +\beta\left\vert
1\right\rangle +\omega^{2}\beta\left\vert 2\right\rangle +\omega
\gamma\left\vert 0\right\rangle +\omega^{2}\gamma\left\vert 1\right\rangle
+\gamma\left\vert 2\right\rangle \right)  .\nonumber
\end{align}
Now the three post-selected states that are sent to Bob, based on the result
obtained by Alice can be determined%
\begin{align*}
\left\vert \phi_{0}\right\rangle  &  =\frac{\alpha+\beta+\gamma}{3}\left\vert
0\right\rangle +\frac{\alpha+\omega^{2}\beta+\omega^{2}\gamma}{3}\left\vert
1\right\rangle +\frac{\alpha+\omega\beta+\omega\gamma}{3}\left\vert
2\right\rangle \\
\left\vert \phi_{1}\right\rangle  &  =\frac{\alpha+\omega^{2}\beta
+\omega\gamma}{3}\left\vert 0\right\rangle +\frac{\alpha+\omega\beta+\gamma
}{3}\left\vert 1\right\rangle +\frac{\alpha+\beta+\omega^{2}\gamma}%
{3}\left\vert 2\right\rangle \\
\left\vert \phi_{2}\right\rangle  &  =\frac{\alpha+\omega\beta+\omega
^{2}\gamma}{3}\left\vert 0\right\rangle +\frac{\alpha+\beta+\omega\gamma}%
{3}\left\vert 1\right\rangle +\frac{\alpha+\omega^{2}\beta+\gamma}%
{3}\left\vert 2\right\rangle .
\end{align*}
To allow Bob unambiguously guess the output of Alice, these three states need
to be perpendicular to each other. To achieve that, they need to be pairwise
perpendicular, which bring us to three equations (perpendicularity of pairs
$\left\vert \phi_{0}\right\rangle \bot\left\vert \phi_{1}\right\rangle $,
$\left\vert \phi_{0}\right\rangle \bot\left\vert \phi_{2}\right\rangle $ and
$\left\vert \phi_{1}\right\rangle \bot\left\vert \phi_{2}\right\rangle $)%
\begin{align}
\left(  \alpha+\beta+\gamma\right)  ^{\ast}\left(  \alpha+\omega^{2}%
\beta+\omega\gamma\right)  +\left(  \alpha+\omega^{2}\beta+\omega^{2}%
\gamma\right)  ^{\ast}\left(  \alpha+\omega\beta+\gamma\right)  +\left(
\alpha+\omega\beta+\omega\gamma\right)  ^{\ast}\left(  \alpha+\beta+\omega
^{2}\gamma\right)   &  =0\label{qutrit_perp}\\
\left(  \alpha+\beta+\gamma\right)  ^{\ast}\left(  \alpha+\omega\beta
+\omega^{2}\gamma\right)  +\left(  \alpha+\omega^{2}\beta+\omega^{2}%
\gamma\right)  ^{\ast}\left(  \alpha+\beta+\omega\gamma\right)  +\left(
\alpha+\omega\beta+\omega\gamma\right)  ^{\ast}\left(  \alpha+\omega^{2}%
\beta+\gamma\right)   &  =0\nonumber\\
\left(  \alpha+\omega^{2}\beta+\omega\gamma\right)  ^{\ast}\left(
\alpha+\omega\beta+\omega^{2}\gamma\right)  +\left(  \alpha+\omega\beta
+\gamma\right)  ^{\ast}\left(  \alpha+\beta+\omega\gamma\right)  +\left(
\alpha+\beta+\omega^{2}\gamma\right)  ^{\ast}\left(  \alpha+\omega^{2}%
\beta+\gamma\right)   &  =0.\nonumber
\end{align}

We can subtract the first two equations of (\ref{qutrit_perp}) and get%
\begin{equation}
\left(  \alpha+\beta+\gamma\right)  ^{\ast}\left(  \omega^{2}-\omega\right)
\left(  \beta-\gamma\right)  +\left(  \alpha+\omega^{2}\beta+\omega^{2}%
\gamma\right)  ^{\ast}\left(  \omega-1\right)  \left(  \beta-\gamma\right)
+\left(  \alpha+\omega\beta+\omega\gamma\right)  ^{\ast}\left(  1-\omega
^{2}\right)  \left(  \beta-\gamma\right)  =0.\label{1-2}%
\end{equation}
This can be further decomposed to
\begin{equation}
\omega\left(  \omega-1\right)  ^{2}\left(  \beta-\gamma\right)  \left(
\beta+\gamma\right)  ^{\ast}=0.\label{qutrit_12_fact}%
\end{equation}
Let us analyze the two possible solutions of (\ref{qutrit_12_fact}).

\subsubsection{$\beta=\gamma$}
If $\beta=\gamma$, from the third equation of (\ref{qutrit_perp}) we get%
\[
\left\vert \alpha+\left(  \omega^{2}+\omega\right)  \beta\right\vert
^{2}+\left\vert \alpha+\left(  1+\omega\right)  \beta\right\vert
^{2}+\left\vert \alpha+\left(  \omega^{2}+1\right)  \beta\right\vert ^{2}=0,
\]
which can obviously only be satisfied for $\alpha=\beta=0$, leading to the
trivial non-normalizable solution $\alpha=\beta=\gamma=0$.

\subsubsection{$\beta=-\gamma$}

For the second solution of (\ref{qutrit_12_fact}), we get from the first
equation of (\ref{qutrit_perp})%
\[
\alpha^{\ast}\left(  \alpha+\omega^{2}\beta-\omega\beta+\alpha+\omega
\beta-\beta+\alpha+\beta-\omega^{2}\beta\right)  =3\left\vert \alpha
\right\vert ^{2}=0.
\]
Plugging this into the third equation of (\ref{qutrit_perp}) yeilds%
\[
\left(  \omega-\omega^{2}\right)^2  \left\vert \beta\right\vert ^{2}+\left(
\omega^{2}-1\right)  \left(  1-\omega^2\right)  \left\vert \beta\right\vert
+\left(  1-\omega\right)  \left(  \omega-1\right)  \left\vert
\beta\right\vert ^{2}=0,
\]
which only has a trivial solution $\beta=0$, leading again to the trivial
non-normalizable solution $\alpha=\beta=\gamma=0$. This concludes the proof
there there does not exist a probe state for Bob that would allow him to learn
with certainty the outcome of the measurement performed by Alice.

\subsection{Parameter counting}

In the previous paragraphs we have shown that unlike the qubit case, for three
unbiased measurements on a qutrit Bob is not able to prepare a probe state that
would allow him to guess the outcome of the measurement performed by Alice with probability $1$.
This however cannot be explained by the simple argument valid for two
measurements on a qutrit, namely that the state obtained by Bob cannot contain
enough information to guess the result. Also in a different perspective
discussed in \cite{Rozpedek2017}, three measurements on a qutrit allow to build up a maximally
entangled state between Alice and Bob and yet it is not sufficient for
reliable guessing. Here we argue by parameter counting that this is a general
feature of the game for any system larger than a qubit, independently on the
number of distinct measurements.

In the most general case, Bob designs a probe state by specifying $2B-2$ real
parameters. He then receives a state of a dimension $A$, which can be viewed
as a mixture of post-selected states specified by the number of $B$ possible
outcomes of the measurement performed by Alice. For successfully learning the
measurement outcome, all these states have to be perpendicular to each other.
This is naturally impossible for $B>A$, simply due to the dimension of the
space, except for special cases discussed below.

Independently on $A$, perpendicularity of $B$ different states can be obtained
by fulfilling $B\left(  B-1\right)  $ simple equations for real parameters.
Interestingly, for a qubit ($B=2$) it holds
\[
B\left(  B-1\right)=2B-2
\]
and thus the number of free parameters is equal to the number of equations to
fulfill. In the previous section we have shown that a solution always exist.

For any $B>2$ it holds
\[
B\left(  B-1\right)  >2B-2.
\]
Hence, for a general situation not all the equations can be fulfilled and thus
Bob cannot guess Alice's outcome with certainty. The exceptions are formed by a
subset of measurements chosen by Alice that are of a measure zero within all
possible set of measurements. The relative cardinality of the subset decreases
with increasing $B$ due to the fact that Bob's choice of the probe state can
only lead to fulfilment of a linear portion from the quadratically increasing
set of conditions and the rest needs to be guaranteed by the selection of
measurements.

\subsection{Special cases}

Let us analyze here, what properties need to be fulfilled for the set of measurements
selected by Alice to allow Bob, at least in principle, to guess with certainty the result
of Alice's measurement.  Let us discuss first the case of $B>A$, which is easier to tackle.

Apparently, by measuring a state of dimension $A$, Bob only can learn up to $A$
different outcomes. Thus he needs to have some a-priori information about the
possible outcomes of Alice's measurement, which have $B>A$ possible outcomes.
Moreover, as we are interested
here in perfect guessing, this a-priori information needs to be complete in a
sense that Bob needs to know about some of the possible outcomes that they do not appear at all --
the other option, knowing with certainty that a result will appear corresponds
to the almost trivial case of all
measurements sharing the same eigenstate.

To achieve this, there must exist a
probe state for which at least $B-A$ measurement results (for each and every measurement) appear with zero
probability. This can only be realized if all measurements share a common
subspace of a dimension of at least $B-A$ by choosing the probe state
perpendicular to this subspace. This naturally leads to a condition that
this subspace must have a dimension smaller than $B$ (there still must exist at least one dimension
outside the subspace in which the probe state can exist).

So we can summarize that the \emph{necessary condition} for perfect
guessing in the case $B>A$ is that there exist a decomposition of the
$B$-dimensional Hilbert space into two non-trivial subspaces such that at least
one of them is of a dimension of at least $B-A$ and for each measurement, each
of its results falls strictly into one of these subspaces.
However, this condition is not \emph{sufficient} -- even if it is
fulfilled, the probe state needs to satisfy also the condition of
mutual perpendicularity of the
post-selected states in the form (\ref{qutrit_perp}).

To check this condition, let us set the dimension of the
Hilbert sub-space containing the probe state to $C$. Then we arrive at $2\left(B-C\right)$ conditions on the probe state
due to the perpendicularity to the rest of the Hilbert space and $2C\left(  C-1\right)  $
conditions for the mutual perpendicularity of the post-selected states.
This leads to in total $2B-4C+2C^2$ conditions, which can
be (potentially) satisfied using $2B-2$ parameters of the probe state only for $C=1$. This however again
leads to the semi-trivial solution of a single shared eigenstate, i.e. that the result of each measurement
is perfectly predetermined.

For the case $B\leq A$ the situation is a bit more complicated. Here Bob gets a state of a
dimension that allows him, at least in principle, to learn enough information for perfect guessing.
This is also the case for qubits, as investigated in the previous section, where Bob can guess in all cases.
This result also can be used to show instances where perfect guessing is possible in higher dimensions.

If we choose $B=2^{n}$, we can view the probe state prepared by Bob as an
$n$-qubit state and prepare it in the form of products of (\ref{psi_Bob}).
We can restrict Alice to use only measurements
factorable on these two-level subspaces, the same for all measurements. But even
this is not enough, the set of measurements shall include all the combinations
of different measurements on different qubits with the same weight (or
appropriate relative weights for the non-uniform extension). For this, rather restricted,
but still non-trivial set of measurements used by Alice, one can use procedures presented
in the previous section to prepare a probe state that shall allow Bob's perfect guessing. This example shows that the
solution based on a single shared eigenstate presented above is not unique for systems larger than a qubit.

\section{Conclusion}\label{chap5}
We have analyzed to what extent the impossibility of guessing an outcome of
one of a prescribed set of measurements is connected with the inaccessibility
of quantum information about the choice of the measurement. Interestingly, we
have shown that specifically for two-dimensional systems (qubits),
availability of coherent quantum information about the choice allows to
correctly guess the outcome of any measurement from any predefined set by
choosing a proper probe-state and final measurement. Both of these choices depend
on the particular form of the measurements in the set, as well as on
frequencies of their appearance in the more complex scenario.

Interestingly, the situation changes dramatically for larger systems. We have
shown explicitly that for three measurements in mutually unbiased bases
performed on a qutrit, perfect guessing is not possible anymore. We have also
shown that this is a general feature for all higher dimensional systems and
sets of measurements for which guessing is possible form a subset of measure
zero among all possible sets.

This result highlights the fundamental difference between the quantum
properties of qubits and higher-dimensional systems, manifested e.g. by
(non-)existence of bound entanglement \cite{Horodecki1998}, or (non-)existence of non-contextual
hidden variable models \cite{Kochen1967}.
Its direct manifestation in the
proposed guessing game motivates further research of direct use of
higher-dimensional states in secure quantum communication, potentially
offering higher level of protection against adversaries.

\section*{Acknowledgments}

We would like to thank Jed Kaniewski and Marcus Huber for fruitful discussions.
This research was supported by the joint Czech-Austrian project MultiQUEST (I 3053-N27 and GF17-33780L), as well as project VEGA 2/0043/15.
MP2 also acknowledges support from Austrian Science Fund (FWF) through the START project Y879-N27.

\bibliography{LOI}{}

\begin{thebibliography}{32}%
\makeatletter
\providecommand \@ifxundefined [1]{%
 \@ifx{#1\undefined}
}%
\providecommand \@ifnum [1]{%
 \ifnum #1\expandafter \@firstoftwo
 \else \expandafter \@secondoftwo
 \fi
}%
\providecommand \@ifx [1]{%
 \ifx #1\expandafter \@firstoftwo
 \else \expandafter \@secondoftwo
 \fi
}%
\providecommand \natexlab [1]{#1}%
\providecommand \enquote  [1]{``#1''}%
\providecommand \bibnamefont  [1]{#1}%
\providecommand \bibfnamefont [1]{#1}%
\providecommand \citenamefont [1]{#1}%
\providecommand \href@noop [0]{\@secondoftwo}%
\providecommand \href [0]{\begingroup \@sanitize@url \@href}%
\providecommand \@href[1]{\@@startlink{#1}\@@href}%
\providecommand \@@href[1]{\endgroup#1\@@endlink}%
\providecommand \@sanitize@url [0]{\catcode `\\12\catcode `\$12\catcode
  `\&12\catcode `\#12\catcode `\^12\catcode `\_12\catcode `\%12\relax}%
\providecommand \@@startlink[1]{}%
\providecommand \@@endlink[0]{}%
\providecommand \url  [0]{\begingroup\@sanitize@url \@url }%
\providecommand \@url [1]{\endgroup\@href {#1}{\urlprefix }}%
\providecommand \urlprefix  [0]{URL }%
\providecommand \Eprint [0]{\href }%
\providecommand \doibase [0]{http://dx.doi.org/}%
\providecommand \selectlanguage [0]{\@gobble}%
\providecommand \bibinfo  [0]{\@secondoftwo}%
\providecommand \bibfield  [0]{\@secondoftwo}%
\providecommand \translation [1]{[#1]}%
\providecommand \BibitemOpen [0]{}%
\providecommand \bibitemStop [0]{}%
\providecommand \bibitemNoStop [0]{.\EOS\space}%
\providecommand \EOS [0]{\spacefactor3000\relax}%
\providecommand \BibitemShut  [1]{\csname bibitem#1\endcsname}%
\let\auto@bib@innerbib\@empty
\bibitem [{\citenamefont {Rozpêdek}\ \emph {et~al.}(2017)\citenamefont
  {Rozpêdek}, \citenamefont {Kaniewski}, \citenamefont {Coles},\ and\
  \citenamefont {Wehner}}]{Rozpedek2017}%
  \BibitemOpen
  \bibfield  {author} {\bibinfo {author} {\bibfnamefont {F.}~\bibnamefont
  {Rozpêdek}}, \bibinfo {author} {\bibfnamefont {J.}~\bibnamefont {Kaniewski}},
  \bibinfo {author} {\bibfnamefont {P.~J.}\ \bibnamefont {Coles}}, \ and\
  \bibinfo {author} {\bibfnamefont {S.}~\bibnamefont {Wehner}},\ }\href
  {http://stacks.iop.org/1367-2630/19/i=2/a=023038} {\bibfield  {journal}
  {\bibinfo  {journal} {New Journal of Physics}\ }\textbf {\bibinfo {volume}
  {19}},\ \bibinfo {pages} {023038} (\bibinfo {year} {2017})}\BibitemShut
  {NoStop}%
\bibitem [{\citenamefont {Nielsen}\ and\ \citenamefont
  {Chuang}(2011)}]{Nielsen:2011:QCQ:1972505}%
  \BibitemOpen
  \bibfield  {author} {\bibinfo {author} {\bibfnamefont {M.~A.}\ \bibnamefont
  {Nielsen}}\ and\ \bibinfo {author} {\bibfnamefont {I.~L.}\ \bibnamefont
  {Chuang}},\ }\href@noop {} {\emph {\bibinfo {title} {Quantum Computation and
  Quantum Information: 10th Anniversary Edition}}},\ \bibinfo {edition} {10th}\
  ed.\ (\bibinfo  {publisher} {Cambridge University Press},\ \bibinfo {address}
  {New York, NY, USA},\ \bibinfo {year} {2011})\BibitemShut {NoStop}%
\bibitem [{\citenamefont {Heisenberg}(1927)}]{Heisenberg1927}%
  \BibitemOpen
  \bibfield  {author} {\bibinfo {author} {\bibfnamefont {W.}~\bibnamefont
  {Heisenberg}},\ }\href {\doibase 10.1007/BF01397280} {\bibfield  {journal}
  {\bibinfo  {journal} {Zeitschrift f{\"u}r Physik}\ }\textbf {\bibinfo
  {volume} {43}},\ \bibinfo {pages} {172} (\bibinfo {year} {1927})}\BibitemShut
  {NoStop}%
\bibitem [{\citenamefont {Kennard}(1927)}]{Kennard1927}%
  \BibitemOpen
  \bibfield  {author} {\bibinfo {author} {\bibfnamefont {E.~H.}\ \bibnamefont
  {Kennard}},\ }\href {\doibase 10.1007/BF01391200} {\bibfield  {journal}
  {\bibinfo  {journal} {Zeitschrift f{\"u}r Physik}\ }\textbf {\bibinfo
  {volume} {44}},\ \bibinfo {pages} {326} (\bibinfo {year} {1927})}\BibitemShut
  {NoStop}%
\bibitem [{\citenamefont {Berta}\ \emph
  {et~al.}(2014{\natexlab{a}})\citenamefont {Berta}, \citenamefont {Coles},\
  and\ \citenamefont {Wehner}}]{PhysRevA.90.062127}%
  \BibitemOpen
  \bibfield  {author} {\bibinfo {author} {\bibfnamefont {M.}~\bibnamefont
  {Berta}}, \bibinfo {author} {\bibfnamefont {P.~J.}\ \bibnamefont {Coles}}, \
  and\ \bibinfo {author} {\bibfnamefont {S.}~\bibnamefont {Wehner}},\ }\href
  {\doibase 10.1103/PhysRevA.90.062127} {\bibfield  {journal} {\bibinfo
  {journal} {Phys. Rev. A}\ }\textbf {\bibinfo {volume} {90}},\ \bibinfo
  {pages} {062127} (\bibinfo {year} {2014}{\natexlab{a}})}\BibitemShut
  {NoStop}%
\bibitem [{\citenamefont {Berta}\ \emph
  {et~al.}(2014{\natexlab{b}})\citenamefont {Berta}, \citenamefont {Fawzi},\
  and\ \citenamefont {Wehner}}]{Berta2014}%
  \BibitemOpen
  \bibfield  {author} {\bibinfo {author} {\bibfnamefont {M.}~\bibnamefont
  {Berta}}, \bibinfo {author} {\bibfnamefont {O.}~\bibnamefont {Fawzi}}, \ and\
  \bibinfo {author} {\bibfnamefont {S.}~\bibnamefont {Wehner}},\ }\href
  {\doibase 10.1109/TIT.2013.2291780} {\bibfield  {journal} {\bibinfo
  {journal} {IEEE Transactions on Information Theory}\ }\textbf {\bibinfo
  {volume} {60}},\ \bibinfo {pages} {1168} (\bibinfo {year}
  {2014}{\natexlab{b}})}\BibitemShut {NoStop}%
\bibitem [{\citenamefont {Christandl}\ and\ \citenamefont
  {Winter}(2005)}]{Christandl2005}%
  \BibitemOpen
  \bibfield  {author} {\bibinfo {author} {\bibfnamefont {M.}~\bibnamefont
  {Christandl}}\ and\ \bibinfo {author} {\bibfnamefont {A.}~\bibnamefont
  {Winter}},\ }\href {\doibase 10.1109/TIT.2005.853338} {\bibfield  {journal}
  {\bibinfo  {journal} {IEEE Transactions on Information Theory}\ }\textbf
  {\bibinfo {volume} {51}},\ \bibinfo {pages} {3159} (\bibinfo {year}
  {2005})}\BibitemShut {NoStop}%
\bibitem [{\citenamefont {Berta}\ \emph {et~al.}(2010)\citenamefont {Berta},
  \citenamefont {Christandl}, \citenamefont {Colbeck}, \citenamefont {Renes},\
  and\ \citenamefont {Renner}}]{Berta2010}%
  \BibitemOpen
  \bibfield  {author} {\bibinfo {author} {\bibfnamefont {M.}~\bibnamefont
  {Berta}}, \bibinfo {author} {\bibfnamefont {M.}~\bibnamefont {Christandl}},
  \bibinfo {author} {\bibfnamefont {R.}~\bibnamefont {Colbeck}}, \bibinfo
  {author} {\bibfnamefont {J.~M.}\ \bibnamefont {Renes}}, \ and\ \bibinfo
  {author} {\bibfnamefont {R.}~\bibnamefont {Renner}},\ }\href {\doibase
  10.1038/nphys1734} {\bibfield  {journal} {\bibinfo  {journal} {Nat Phys}\
  }\textbf {\bibinfo {volume} {6}},\ \bibinfo {pages} {659} (\bibinfo {year}
  {2010})}\BibitemShut {NoStop}%
\bibitem [{\citenamefont {Coles}\ \emph {et~al.}(2012)\citenamefont {Coles},
  \citenamefont {Colbeck}, \citenamefont {Yu},\ and\ \citenamefont
  {Zwolak}}]{Coles2012}%
  \BibitemOpen
  \bibfield  {author} {\bibinfo {author} {\bibfnamefont {P.~J.}\ \bibnamefont
  {Coles}}, \bibinfo {author} {\bibfnamefont {R.}~\bibnamefont {Colbeck}},
  \bibinfo {author} {\bibfnamefont {L.}~\bibnamefont {Yu}}, \ and\ \bibinfo
  {author} {\bibfnamefont {M.}~\bibnamefont {Zwolak}},\ }\href {\doibase
  10.1103/PhysRevLett.108.210405} {\bibfield  {journal} {\bibinfo  {journal}
  {Phys. Rev. Lett.}\ }\textbf {\bibinfo {volume} {108}},\ \bibinfo {pages}
  {210405} (\bibinfo {year} {2012})}\BibitemShut {NoStop}%
\bibitem [{\citenamefont {Coles}\ \emph {et~al.}(2011)\citenamefont {Coles},
  \citenamefont {Yu}, \citenamefont {Gheorghiu},\ and\ \citenamefont
  {Griffiths}}]{Coles2011}%
  \BibitemOpen
  \bibfield  {author} {\bibinfo {author} {\bibfnamefont {P.~J.}\ \bibnamefont
  {Coles}}, \bibinfo {author} {\bibfnamefont {L.}~\bibnamefont {Yu}}, \bibinfo
  {author} {\bibfnamefont {V.}~\bibnamefont {Gheorghiu}}, \ and\ \bibinfo
  {author} {\bibfnamefont {R.~B.}\ \bibnamefont {Griffiths}},\ }\href {\doibase
  10.1103/PhysRevA.83.062338} {\bibfield  {journal} {\bibinfo  {journal} {Phys.
  Rev. A}\ }\textbf {\bibinfo {volume} {83}},\ \bibinfo {pages} {062338}
  (\bibinfo {year} {2011})}\BibitemShut {NoStop}%
\bibitem [{\citenamefont {Dupuis}\ \emph {et~al.}(2015)\citenamefont {Dupuis},
  \citenamefont {Fawzi},\ and\ \citenamefont {Wehner}}]{Dupuis2015}%
  \BibitemOpen
  \bibfield  {author} {\bibinfo {author} {\bibfnamefont {F.}~\bibnamefont
  {Dupuis}}, \bibinfo {author} {\bibfnamefont {O.}~\bibnamefont {Fawzi}}, \
  and\ \bibinfo {author} {\bibfnamefont {S.}~\bibnamefont {Wehner}},\ }\href
  {\doibase 10.1109/TIT.2014.2371464} {\bibfield  {journal} {\bibinfo
  {journal} {IEEE Transactions on Information Theory}\ }\textbf {\bibinfo
  {volume} {61}},\ \bibinfo {pages} {1093} (\bibinfo {year}
  {2015})}\BibitemShut {NoStop}%
\bibitem [{\citenamefont {Frank}\ and\ \citenamefont {Lieb}(2013)}]{Frank2013}%
  \BibitemOpen
  \bibfield  {author} {\bibinfo {author} {\bibfnamefont {R.~L.}\ \bibnamefont
  {Frank}}\ and\ \bibinfo {author} {\bibfnamefont {E.~H.}\ \bibnamefont
  {Lieb}},\ }\href {\doibase 10.1007/s00220-013-1775-1} {\bibfield  {journal}
  {\bibinfo  {journal} {Communications in Mathematical Physics}\ }\textbf
  {\bibinfo {volume} {323}},\ \bibinfo {pages} {487} (\bibinfo {year}
  {2013})}\BibitemShut {NoStop}%
\bibitem [{\citenamefont {Furrer}\ \emph {et~al.}(2014)\citenamefont {Furrer},
  \citenamefont {Berta}, \citenamefont {Tomamichel}, \citenamefont {Scholz},\
  and\ \citenamefont {Christandl}}]{Furrer2014}%
  \BibitemOpen
  \bibfield  {author} {\bibinfo {author} {\bibfnamefont {F.}~\bibnamefont
  {Furrer}}, \bibinfo {author} {\bibfnamefont {M.}~\bibnamefont {Berta}},
  \bibinfo {author} {\bibfnamefont {M.}~\bibnamefont {Tomamichel}}, \bibinfo
  {author} {\bibfnamefont {V.~B.}\ \bibnamefont {Scholz}}, \ and\ \bibinfo
  {author} {\bibfnamefont {M.}~\bibnamefont {Christandl}},\ }\href {\doibase
  http://dx.doi.org/10.1063/1.4903989} {\bibfield  {journal} {\bibinfo
  {journal} {Journal of Mathematical Physics}\ }\textbf {\bibinfo {volume}
  {55}},\ \bibinfo {pages} {122205} (\bibinfo {year} {2014})}\BibitemShut
  {NoStop}%
\bibitem [{\citenamefont {Hall}(1995)}]{Hall1995}%
  \BibitemOpen
  \bibfield  {author} {\bibinfo {author} {\bibfnamefont {M.~J.~W.}\
  \bibnamefont {Hall}},\ }\href {\doibase 10.1103/PhysRevLett.74.3307}
  {\bibfield  {journal} {\bibinfo  {journal} {Phys. Rev. Lett.}\ }\textbf
  {\bibinfo {volume} {74}},\ \bibinfo {pages} {3307} (\bibinfo {year}
  {1995})}\BibitemShut {NoStop}%
\bibitem [{\citenamefont {Liu}\ \emph {et~al.}(2015)\citenamefont {Liu},
  \citenamefont {Mu},\ and\ \citenamefont {Fan}}]{Liu2015}%
  \BibitemOpen
  \bibfield  {author} {\bibinfo {author} {\bibfnamefont {S.}~\bibnamefont
  {Liu}}, \bibinfo {author} {\bibfnamefont {L.-Z.}\ \bibnamefont {Mu}}, \ and\
  \bibinfo {author} {\bibfnamefont {H.}~\bibnamefont {Fan}},\ }\href {\doibase
  10.1103/PhysRevA.91.042133} {\bibfield  {journal} {\bibinfo  {journal} {Phys.
  Rev. A}\ }\textbf {\bibinfo {volume} {91}},\ \bibinfo {pages} {042133}
  (\bibinfo {year} {2015})}\BibitemShut {NoStop}%
\bibitem [{\citenamefont {Korzekwa}\ \emph {et~al.}(2014)\citenamefont
  {Korzekwa}, \citenamefont {Lostaglio}, \citenamefont {Jennings},\ and\
  \citenamefont {Rudolph}}]{Korzekwa2014}%
  \BibitemOpen
  \bibfield  {author} {\bibinfo {author} {\bibfnamefont {K.}~\bibnamefont
  {Korzekwa}}, \bibinfo {author} {\bibfnamefont {M.}~\bibnamefont {Lostaglio}},
  \bibinfo {author} {\bibfnamefont {D.}~\bibnamefont {Jennings}}, \ and\
  \bibinfo {author} {\bibfnamefont {T.}~\bibnamefont {Rudolph}},\ }\href
  {\doibase 10.1103/PhysRevA.89.042122} {\bibfield  {journal} {\bibinfo
  {journal} {Phys. Rev. A}\ }\textbf {\bibinfo {volume} {89}},\ \bibinfo
  {pages} {042122} (\bibinfo {year} {2014})}\BibitemShut {NoStop}%
\bibitem [{\citenamefont {Luo}(2005)}]{Luo2005}%
  \BibitemOpen
  \bibfield  {author} {\bibinfo {author} {\bibfnamefont {S.~L.}\ \bibnamefont
  {Luo}},\ }\href {\doibase 10.1007/s11232-005-0098-6} {\bibfield  {journal}
  {\bibinfo  {journal} {Theoretical and Mathematical Physics}\ }\textbf
  {\bibinfo {volume} {143}},\ \bibinfo {pages} {681} (\bibinfo {year}
  {2005})}\BibitemShut {NoStop}%
\bibitem [{\citenamefont {Renes}\ and\ \citenamefont
  {Boileau}(2009)}]{Renes2009}%
  \BibitemOpen
  \bibfield  {author} {\bibinfo {author} {\bibfnamefont {J.~M.}\ \bibnamefont
  {Renes}}\ and\ \bibinfo {author} {\bibfnamefont {J.-C.}\ \bibnamefont
  {Boileau}},\ }\href {\doibase 10.1103/PhysRevLett.103.020402} {\bibfield
  {journal} {\bibinfo  {journal} {Phys. Rev. Lett.}\ }\textbf {\bibinfo
  {volume} {103}},\ \bibinfo {pages} {020402} (\bibinfo {year}
  {2009})}\BibitemShut {NoStop}%
\bibitem [{\citenamefont {Sánchez-Ruiz}(1995)}]{Sanchez-Ruiz1995}%
  \BibitemOpen
  \bibfield  {author} {\bibinfo {author} {\bibfnamefont {J.}~\bibnamefont
  {Sánchez-Ruiz}},\ }\href {\doibase
  http://dx.doi.org/10.1016/0375-9601(95)00219-S} {\bibfield  {journal}
  {\bibinfo  {journal} {Physics Letters A}\ }\textbf {\bibinfo {volume}
  {201}},\ \bibinfo {pages} {125 } (\bibinfo {year} {1995})}\BibitemShut
  {NoStop}%
\bibitem [{\citenamefont {Ludwig}(1964)}]{Ludwig1964}%
  \BibitemOpen
  \bibfield  {author} {\bibinfo {author} {\bibfnamefont {G.}~\bibnamefont
  {Ludwig}},\ }\href@noop {} {\bibfield  {journal} {\bibinfo  {journal} {Z.
  Physik}\ }\textbf {\bibinfo {volume} {181}},\ \bibinfo {pages} {233}
  (\bibinfo {year} {1964})}\BibitemShut {NoStop}%
\bibitem [{\citenamefont {Busch}\ and\ \citenamefont
  {Lahti}(1984)}]{Busch1984}%
  \BibitemOpen
  \bibfield  {author} {\bibinfo {author} {\bibfnamefont {P.}~\bibnamefont
  {Busch}}\ and\ \bibinfo {author} {\bibfnamefont {P.~J.}\ \bibnamefont
  {Lahti}},\ }\href {\doibase 10.1103/PhysRevD.29.1634} {\bibfield  {journal}
  {\bibinfo  {journal} {Phys. Rev. D}\ }\textbf {\bibinfo {volume} {29}},\
  \bibinfo {pages} {1634} (\bibinfo {year} {1984})}\BibitemShut {NoStop}%
\bibitem [{\citenamefont {Busch}(1986)}]{Busch1986}%
  \BibitemOpen
  \bibfield  {author} {\bibinfo {author} {\bibfnamefont {P.}~\bibnamefont
  {Busch}},\ }\href {\doibase 10.1103/PhysRevD.33.2253} {\bibfield  {journal}
  {\bibinfo  {journal} {Phys. Rev. D}\ }\textbf {\bibinfo {volume} {33}},\
  \bibinfo {pages} {2253} (\bibinfo {year} {1986})}\BibitemShut {NoStop}%
\bibitem [{\citenamefont {de~Muynck}\ and\ \citenamefont
  {Martens}(1990)}]{Muynck1990}%
  \BibitemOpen
  \bibfield  {author} {\bibinfo {author} {\bibfnamefont {W.~M.}\ \bibnamefont
  {de~Muynck}}\ and\ \bibinfo {author} {\bibfnamefont {H.}~\bibnamefont
  {Martens}},\ }\href {\doibase 10.1103/PhysRevA.42.5079} {\bibfield  {journal}
  {\bibinfo  {journal} {Phys. Rev. A}\ }\textbf {\bibinfo {volume} {42}},\
  \bibinfo {pages} {5079} (\bibinfo {year} {1990})}\BibitemShut {NoStop}%
\bibitem [{\citenamefont {Lahti}(2003)}]{Lahti2003}%
  \BibitemOpen
  \bibfield  {author} {\bibinfo {author} {\bibfnamefont {P.}~\bibnamefont
  {Lahti}},\ }\href {\doibase 10.1023/A:1025406103210} {\bibfield  {journal}
  {\bibinfo  {journal} {International Journal of Theoretical Physics}\ }\textbf
  {\bibinfo {volume} {42}},\ \bibinfo {pages} {893} (\bibinfo {year}
  {2003})}\BibitemShut {NoStop}%
\bibitem [{\citenamefont {Wolf}\ \emph {et~al.}(2009)\citenamefont {Wolf},
  \citenamefont {Perez-Garcia},\ and\ \citenamefont {Fernandez}}]{Wolf2009}%
  \BibitemOpen
  \bibfield  {author} {\bibinfo {author} {\bibfnamefont {M.~M.}\ \bibnamefont
  {Wolf}}, \bibinfo {author} {\bibfnamefont {D.}~\bibnamefont {Perez-Garcia}},
  \ and\ \bibinfo {author} {\bibfnamefont {C.}~\bibnamefont {Fernandez}},\
  }\href {\doibase 10.1103/PhysRevLett.103.230402} {\bibfield  {journal}
  {\bibinfo  {journal} {Phys. Rev. Lett.}\ }\textbf {\bibinfo {volume} {103}},\
  \bibinfo {pages} {230402} (\bibinfo {year} {2009})}\BibitemShut {NoStop}%
\bibitem [{\citenamefont {Reeb}\ \emph {et~al.}(2013)\citenamefont {Reeb},
  \citenamefont {Reitzner},\ and\ \citenamefont {Wolf}}]{Reeb2013}%
  \BibitemOpen
  \bibfield  {author} {\bibinfo {author} {\bibfnamefont {D.}~\bibnamefont
  {Reeb}}, \bibinfo {author} {\bibfnamefont {D.}~\bibnamefont {Reitzner}}, \
  and\ \bibinfo {author} {\bibfnamefont {M.~M.}\ \bibnamefont {Wolf}},\ }\href
  {http://stacks.iop.org/1751-8121/46/i=46/a=462002} {\bibfield  {journal}
  {\bibinfo  {journal} {Journal of Physics A: Mathematical and Theoretical}\
  }\textbf {\bibinfo {volume} {46}},\ \bibinfo {pages} {462002} (\bibinfo
  {year} {2013})}\BibitemShut {NoStop}%
\bibitem [{\citenamefont {Uola}\ \emph {et~al.}(2014)\citenamefont {Uola},
  \citenamefont {Moroder},\ and\ \citenamefont {G\"uhne}}]{Uola2014}%
  \BibitemOpen
  \bibfield  {author} {\bibinfo {author} {\bibfnamefont {R.}~\bibnamefont
  {Uola}}, \bibinfo {author} {\bibfnamefont {T.}~\bibnamefont {Moroder}}, \
  and\ \bibinfo {author} {\bibfnamefont {O.}~\bibnamefont {G\"uhne}},\ }\href
  {\doibase 10.1103/PhysRevLett.113.160403} {\bibfield  {journal} {\bibinfo
  {journal} {Phys. Rev. Lett.}\ }\textbf {\bibinfo {volume} {113}},\ \bibinfo
  {pages} {160403} (\bibinfo {year} {2014})}\BibitemShut {NoStop}%
\bibitem [{\citenamefont {Heinosaari}\ \emph {et~al.}(2014)\citenamefont
  {Heinosaari}, \citenamefont {Miyadera},\ and\ \citenamefont
  {Reitzner}}]{Heinosaari2014}%
  \BibitemOpen
  \bibfield  {author} {\bibinfo {author} {\bibfnamefont {T.}~\bibnamefont
  {Heinosaari}}, \bibinfo {author} {\bibfnamefont {T.}~\bibnamefont
  {Miyadera}}, \ and\ \bibinfo {author} {\bibfnamefont {D.}~\bibnamefont
  {Reitzner}},\ }\href {\doibase 10.1007/s10701-013-9761-1} {\bibfield
  {journal} {\bibinfo  {journal} {Foundations of Physics}\ }\textbf {\bibinfo
  {volume} {44}},\ \bibinfo {pages} {34} (\bibinfo {year} {2014})}\BibitemShut
  {NoStop}%
\bibitem [{\citenamefont {Heinosaari}\ \emph {et~al.}(2015)\citenamefont
  {Heinosaari}, \citenamefont {Kiukas},\ and\ \citenamefont
  {Reitzner}}]{Heinosaari2015}%
  \BibitemOpen
  \bibfield  {author} {\bibinfo {author} {\bibfnamefont {T.}~\bibnamefont
  {Heinosaari}}, \bibinfo {author} {\bibfnamefont {J.}~\bibnamefont {Kiukas}},
  \ and\ \bibinfo {author} {\bibfnamefont {D.}~\bibnamefont {Reitzner}},\
  }\href {\doibase 10.1103/PhysRevA.92.022115} {\bibfield  {journal} {\bibinfo
  {journal} {Phys. Rev. A}\ }\textbf {\bibinfo {volume} {92}},\ \bibinfo
  {pages} {022115} (\bibinfo {year} {2015})}\BibitemShut {NoStop}%
\bibitem [{\citenamefont {Wootters}\ and\ \citenamefont
  {Fields}(1989)}]{WOOTTERS1989363}%
  \BibitemOpen
  \bibfield  {author} {\bibinfo {author} {\bibfnamefont {W.}~\bibnamefont
  {Wootters}}\ and\ \bibinfo {author} {\bibfnamefont {B.~D.}\ \bibnamefont
  {Fields}},\ }\href {\doibase http://dx.doi.org/10.1016/0003-4916(89)90322-9}
  {\bibfield  {journal} {\bibinfo  {journal} {Annals of Physics}\ }\textbf
  {\bibinfo {volume} {191}},\ \bibinfo {pages} {363 } (\bibinfo {year}
  {1989})}\BibitemShut {NoStop}%
\bibitem [{\citenamefont {Horodecki}\ \emph {et~al.}(1998)\citenamefont
  {Horodecki}, \citenamefont {Horodecki},\ and\ \citenamefont
  {Horodecki}}]{Horodecki1998}%
  \BibitemOpen
  \bibfield  {author} {\bibinfo {author} {\bibfnamefont {M.}~\bibnamefont
  {Horodecki}}, \bibinfo {author} {\bibfnamefont {P.}~\bibnamefont
  {Horodecki}}, \ and\ \bibinfo {author} {\bibfnamefont {R.}~\bibnamefont
  {Horodecki}},\ }\href {\doibase 10.1103/PhysRevLett.80.5239} {\bibfield
  {journal} {\bibinfo  {journal} {Phys. Rev. Lett.}\ }\textbf {\bibinfo
  {volume} {80}},\ \bibinfo {pages} {5239} (\bibinfo {year}
  {1998})}\BibitemShut {NoStop}%
\bibitem [{\citenamefont {Kochen}\ and\ \citenamefont
  {Specker}(1967)}]{Kochen1967}%
  \BibitemOpen
  \bibfield  {author} {\bibinfo {author} {\bibfnamefont {S.}~\bibnamefont
  {Kochen}}\ and\ \bibinfo {author} {\bibfnamefont {E.~P.}\ \bibnamefont
  {Specker}},\ }\href@noop {} {\bibfield  {journal} {\bibinfo  {journal}
  {Journal of Mathematics and Mechanics}\ }\textbf {\bibinfo {volume} {17}},\
  \bibinfo {pages} {59} (\bibinfo {year} {1967})}\BibitemShut {NoStop}%
\end{thebibliography}%

\end{document}